\documentclass[12pt]{article}
\usepackage[utf8]{inputenc}
\usepackage[T1]{fontenc}
\usepackage[margin=1.1in]{geometry}
\usepackage{setspace}
\usepackage{microtype}
\usepackage{graphicx}
\usepackage{amsmath}
\usepackage{hyperref}
\usepackage{listings}
\usepackage{booktabs}
\usepackage[affil-it]{authblk} 

\usepackage{url}
\usepackage[numbers,sort&compress]{natbib}

\title{The State of Papers, Retractions, and Preprints: Evidence from the CrossRef Database (2004–2024)}

\author[1,*]{Khalid M. Saqr}

\affil{College of Engineering and Technology, AASTMT, 1029 Alexandria, Egypt}

\affil[*]{Corresponding author: $k.saqr@aast.edu$}

\date{}

\begin{document}

\maketitle

\begin{abstract}

A 20-year analysis of CrossRef metadata demonstrates that global scholarly output—encompassing publications, retractions, and preprints—exhibits strikingly inertial growth, well-described by exponential, quadratic, and logistic models with nearly indistinguishable goodness-of-fit. Retraction dynamics, in particular, remain stable and minimally affected by the COVID-19 shock, which contributed less than 1\% to total notices. Since 2004, publications doubled every 9.8 years, retractions every 11.4 years, and preprints at the fastest rate, every 5.6 years. The findings underscore a system primed for ongoing stress at unchanged structural bottlenecks. Although model forecasts diverge beyond 2024, the evidence suggests that the future trajectory of scholarly communication will be determined by persistent systemic inertia rather than episodic disruptions—unless intentionally redirected by policy or AI-driven reform.

\end{abstract}

\section*{Introduction}

The past two decades have witnessed an unprecedented escalation in scholarly output, propelled by digital infrastructures, preprint repositories, and global crises such as COVID-19. The peer review system, long considered the gatekeeper of scientific integrity, now faces acute strain. Multiple recent studies have documented this “peer reviewer crisis,” finding that editors must issue more invitations than ever before to secure reviewers, driving up publication delays and threatening the sustainability of traditional editorial workflows\citep{Tropini2023,Fernandez-Llimos2019,Parrish2022619,Parrish2024171}. Despite improvements in discovery and communication technologies, these systemic bottlenecks have persisted and even worsened, with ripple effects across the research ecosystem.

Large-scale, data-driven analyses have begun to map this transformation in detail. Davidson et al.\citep{Davidson2024} found that preprints and peer-reviewed COVID-19 trial articles largely report consistent effect estimates, supporting the reliability of rapid-distribution models, though some differences in outcome reporting persist. Whitaker et al.\citep{Whitaker2025} showed that, during the pandemic, publication dynamics and citation behaviors changed fundamentally: preprints surged, their prominence fluctuated, and both public and scientific attention focused more tightly on COVID-19 research. Tsunoda et al.\citep{Tsunoda202557} demonstrated that preprint-disseminated articles, once published, receive substantially higher citation counts than directly submitted journal articles, confirming the audience-amplifying effect of preprints in periods of crisis and rapid change.

Yet, a striking feature of this ecosystem is its inertia: even under the duress of pandemic-scale shocks, the core synchrony between scientific growth and error-correction remains. Previous meta-research often relied on a single regression form (typically exponential), potentially missing inflection points and nonlinearities that signal real-world constraints. There remains a critical need to test, side-by-side, multiple growth models—especially as the system transitions from expansion to saturation.

Here, we address this by leveraging comprehensive CrossRef metadata to model annual trends in publications, retractions, and preprints from 2004 to 2024, systematically comparing exponential, quadratic, and logistic fits. By quantifying both the inertia and adaptability of the science system, we identify the boundaries of current growth, the resilience of error-correction mechanisms, and the early warning signs of a new structural era in scholarly communication.

\section*{Results}

\begin{figure}[ht]
  \centering
  \includegraphics[width=0.9\textwidth]{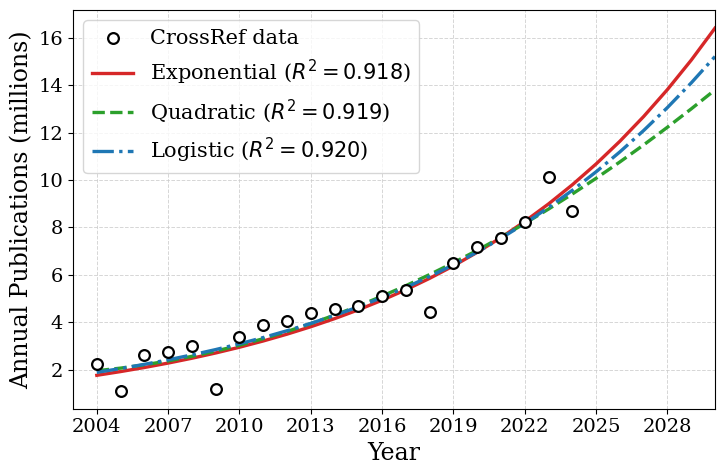}
  \caption{Annual CrossRef-indexed research publications (2004--2024, circles) and forward projection (to 2030) compared with exponential (red), quadratic (green dashed), and logistic (blue dash-dotted) model fits. All models closely match past data; only forward trends diverge.}
  \label{fig:pubs}
\end{figure}

Between 2004 and 2024, the annual number of CrossRef-indexed scholarly publications increased more than fourfold, rising from 2.28 million to 10.12 million. All three model types—logistic ($R^2 = 0.920$), quadratic ($R^2 = 0.919$), and exponential ($R^2 = 0.918$)—fit the historical trajectory with near-indistinguishable precision (Fig.~\ref{fig:pubs}). Forward projections to 2030, however, show clear separation: only the logistic model anticipates saturation, while exponential and quadratic fits forecast unbounded acceleration. These results reveal that, as of 2024, the scholarly system is still governed by inertial momentum, and major shifts in model regime have yet to materialize.

\subsection*{Retraction Trends}

\begin{figure}[ht]
  \centering
  \includegraphics[width=0.9\textwidth]{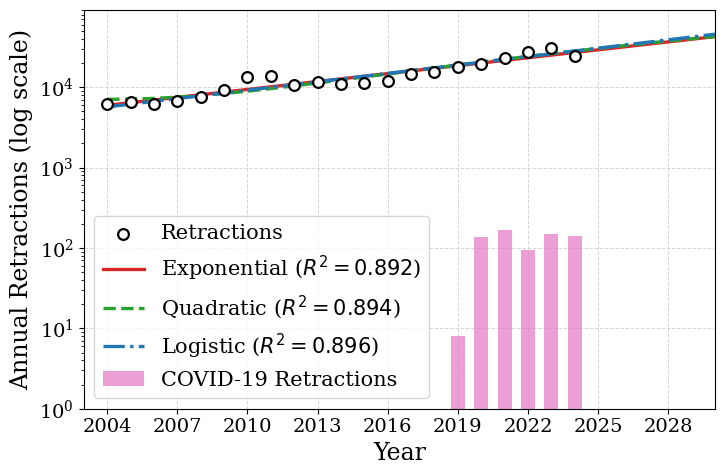}
  \caption{Retractions (2004--2024, circles) with COVID-19-related retractions (bars, 2019--2024). Model fits and forward projections (to 2030) are shown for exponential, quadratic, and logistic functions.}
  \label{fig:rets}
\end{figure}

Retraction notices grew from 6,099 to 30,689. The fits are nearly identical: $R^2 = 0.896$ (logistic), $0.894$ (quadratic), $0.892$ (exponential). Despite the COVID-19 pandemic and sporadic metadata correction spikes (2010–11, 2023), the trend remained inertially smooth (Fig.~\ref{fig:rets}). COVID-19 retractions, which peaked at 165 and never exceeded 1\% of annual totals, represent a shock at least two orders of magnitude below the main trend. The resilience of the main time series—and the close matching of all model fits—demonstrates that retraction growth is best understood as an inertial process, relatively impervious to external disturbances.

\subsection*{Preprints Hypergrowth and Regime Transition}

\begin{figure}[ht]
  \centering
  \includegraphics[width=0.9\textwidth]{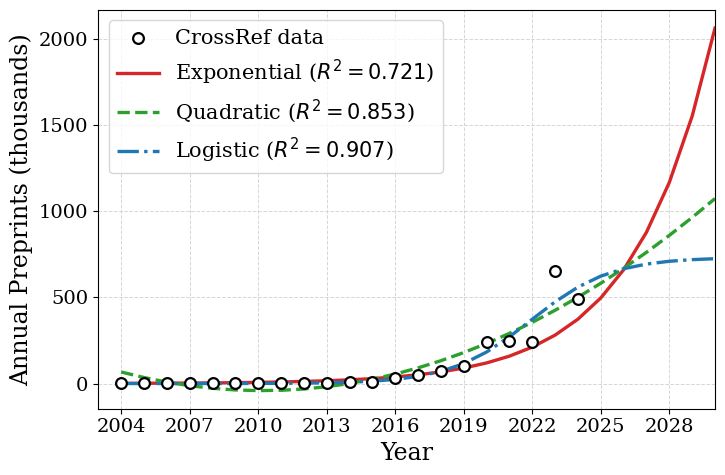}
  \caption{Annual CrossRef-indexed preprints (2004--2024, circles) with fitted exponential, quadratic, and logistic models and forward projections to 2030.}
  \label{fig:preprints}
\end{figure}

Preprints show the fastest doubling (5.6 years) and steeper nonlinearity. The logistic fit ($R^2=0.907$) slightly outperforms the quadratic ($0.853$) and greatly exceeds the exponential ($0.721$), yet all fits align closely through 2024 (Fig.~\ref{fig:preprints}). The convergence in future forecasts (post-2024) indicates a regime where the new wave of preprint servers may outpace legacy controls—especially as generative AI accelerates production and peer review lags.

\subsection*{Inter-series Correlation Analysis}

To evaluate the degree of statistical association between scholarly publication, retraction, and preprint trends, we computed Pearson’s correlation coefficient ($r$), Spearman’s rank correlation coefficient ($\rho$), and Kendall’s tau ($\tau$) for all pairs of time series. Correlations were calculated both for the full annual time series (2004--2024) and for the retraction series with COVID-19-related retractions subtracted, to control for pandemic-specific shocks.

Table~\ref{tab:correlations} summarizes the results. The publication-preprint pair exhibited extremely strong positive correlations across all measures (Pearson $r=0.992$, Spearman $\rho=0.974$, Kendall $\tau=0.882$, all $p<0.001$), indicating synchronized exponential growth and tightly coupled expansion dynamics. The correlation between publications and total retractions was moderate (Pearson $r=0.855$), but dropped further for the rank-based metrics (Spearman $\rho=0.779$, Kendall $\tau=0.623$), reflecting nonlinear and episodic trends (e.g., metadata corrections). Removing COVID-19 retractions had negligible impact on these coefficients, reinforcing the finding that pandemic-era events were not a major driver of systemic synchrony. Preprints and retractions showed the lowest, yet still positive, correlations (Pearson $r=0.742$), consistent with preprints operating as a partially independent subsystem.


\begin{table}[ht]
    \centering
    \caption{\textbf{Statistical correlations (Pearson $r$, Spearman $\rho$, Kendall $\tau$) and $p$-values between key CrossRef time series (2004--2024).} ``Retractions*'' denotes retractions minus COVID-19-related items.}
    \small
    \begin{tabular}{@{}lcccccc@{}}
        \toprule
        \textbf{Series Pair} & \textbf{Pearson $r$} & \textbf{$p$} & \textbf{Spearman $\rho$} & \textbf{$p$} & \textbf{Kendall $\tau$} & \textbf{$p$} \\
        \midrule
        Publications--Preprints       & 0.992 & $<$0.001 & 0.974 & $<$0.001 & 0.882 & $<$0.001 \\
        Publications--Retractions     & 0.855 & $<$0.001 & 0.779 & $<$0.001 & 0.623 & $<$0.001 \\
        Publications--Retractions*    & 0.851 & $<$0.001 & 0.776 & $<$0.001 & 0.618 & $<$0.001 \\
        Preprints--Retractions        & 0.742 & 0.001    & 0.699 & 0.002    & 0.547 & 0.003    \\
        \bottomrule
    \end{tabular}
    \label{tab:correlations}
\end{table}

\section*{Discussion}

Table~\ref{tab:doubling} summarizes the close fit of all three models and the system’s inertial doubling times. The fact that retractions, publications, and preprints all match logistic, quadratic, and exponential models over two decades, despite major global disruptions, is strong evidence for deep systemic inertia. The COVID-19 shock—barely visible in the aggregate—demonstrates that episodic events alone are insufficient to shift long-term trends.

\begin{table}[ht]
    \centering
    \caption{Doubling times and model fits for annual CrossRef publication, retraction, and preprint series, 2004--2024.}
    \label{tab:doubling}
    \begin{tabular}{lccc}
        \hline
        & Publications & Retractions & Preprints \\
        \hline
        Fitted Model   & Logistic    & Logistic    & Logistic    \\
        $R^2$ (logistic) & 0.920     & 0.896       & 0.907       \\
        Doubling Time (years) & 9.8 & 11.4 & 5.6 \\
        \hline
    \end{tabular}
\end{table}

This inertial behavior is both opportunity and warning. On one hand, it means that the research ecosystem is robust: even dramatic increases in output or temporary errors do not immediately destabilize its integrity mechanisms. On the other hand, the inertia also signals risk: as the adoption of generative AI further accelerates publication, the error-correction and retraction processes will likely continue to lag, leading to a growing "residue" of undetected error or misconduct unless active interventions are deployed.

The close matching of multiple regression models also suggests that, for at least the next cycle, forecasts of future trends—whether of growth or crisis—should be understood as probabilistic “inertia cones”: unless a true structural innovation occurs, the current regime will simply continue forward, constrained only by its own doubling time. The onset of AI-driven writing, reviewing, and error-detection could be the catalyst for a new transition, but as of 2024, this remains a horizon scenario.

The strong and statistically significant correlations among publications, retractions, and preprints—unchanged even after removal of COVID-19-related retractions—suggest that these dimensions of the scholarly ecosystem are governed by shared structural and systemic forces. This “inertial synchrony” is reflected in the near-equivalent $R^2$ values achieved by different regression models (logistic, quadratic, and exponential) over the historical period, and is further supported by the finding that even major shocks such as the COVID-19 pandemic exerted only transient, localized effects rather than fundamentally reordering time series relationships.

These results strengthen the case for interpreting the growth of retractions as primarily an inertial, system-driven phenomenon—tightly linked to expansion in output, but subject to its own lag and plateau dynamics. The synchrony also highlights the limits of “shock therapy” interventions, such as those observed during the pandemic aiming at accelerating science to meet regulatory goals, which failed to disrupt the overall trajectory or alter underlying correction rates. The observed doubling times and plateau forecasts should thus be viewed not merely as statistical projections, but as evidence for the persistence of deeply embedded, mutually reinforcing feedbacks within the science system.

Going forward, this synchrony—and its resistance to exogenous disruption—implies that meaningful improvements in research integrity and error correction will require interventions that are systemic in nature, capable of shifting the “inertial” trajectory of the entire scholarly infrastructure, rather than isolated responses to periodic crises.

Scientific governance should therefore focus on identifying inflection points before they appear in time series data, proactively bridging the lag between output and quality control, and preparing for a possible regime shift triggered by AI or policy innovation. Until then, the primary law remains inertia—both in progress and in error.

\section*{Methods}

\subsection*{Data Acquisition via CrossRef REST API}

All data in this study were directly compiled from the official CrossRef REST API (\url{https://api.crossref.org/works}), which provides authoritative, up-to-date metadata on scholarly outputs registered with CrossRef member publishers. To ensure comprehensive coverage and reproducibility, queries were issued annually for four categories: (1) total publications, (2) retractions, (3) COVID-19-related retractions, and (4) preprints. For each year from 2004 to 2024, total publications were retrieved using date-range filters on the \texttt{from-pub-date} and \texttt{until-pub-date} parameters. Retractions were extracted by combining \texttt{update-type:retraction} filters with characteristic title strings (“Retraction”, “Retracted”, “Retraction Notice”, etc.), while COVID-19 retractions additionally required title terms such as “COVID-19”, “SARS-CoV-2”, or “Coronavirus”. Preprints were isolated using the \texttt{type:posted-content} filter. Each query returned a JSON object, from which the annual count was extracted from the \texttt{message.total-results} field. The computer code to reproduce the results from the CrossRef database is available on Github (see Code Availability).

\subsection*{Implementation and Script Architecture}

Data harvesting was automated using the open-source \texttt{CrossRef Fetcher} Python script (available on GitHub, see Code and Data Availability). The script establishes a persistent \texttt{requests.Session} with robust retry logic via \texttt{urllib3.util.Retry} to mitigate server errors and rate limits, and logs all timing and failure events to maximize completeness and auditability. For each data type and year, the script assembles the appropriate REST API query parameters and issues the request, automatically retrying as necessary. All outputs are indexed by year and saved to CSV files.

\subsection*{Data Output Structure}

The primary output is a time series CSV file (\texttt{analysis\_time\_series.csv}) containing the year-wise counts for each category: \texttt{Publications}, \texttt{Retractions}, \texttt{COVID\_Retraction}, and \texttt{Preprints}. Additional outputs include \texttt{regression\_fits\_detailed.csv}, which contains fitted values for all regression models across the time series, and \texttt{regression\_summary.csv}, which summarizes the optimal parameter values, $R^2$ scores, and doubling times for each fit. Correlation analysis results are stored in \texttt{correlations\_summary.csv}, reporting Pearson, Spearman, and Kendall coefficients and associated $p$-values for all pairwise comparisons.

\subsection*{Regression Models and Statistical Analysis}

Each time series was fit to three canonical growth models: exponential, quadratic, and logistic. The exponential model follows $y = ae^{bx}$, the quadratic model $y = ax^2 + bx + c$, and the logistic model $y = \frac{K}{1 + e^{-b(x - x_0)}}$. Nonlinear fitting was performed using SciPy’s \texttt{curve\_fit} and \texttt{linregress} routines, and the coefficient of determination ($R^2$) was calculated for each fit. For each series, the doubling time $T_2$ was computed analytically from the best-fit parameters.

Model selection was based on comparative $R^2$ and visual plausibility, particularly regarding forward projections (through 2030). All regression parameters, fits, and $R^2$ values are reported in \texttt{regression\_summary.csv} and discussed in the Results.

\subsection*{Correlation Analysis}

To assess the interdependence of publication, retraction, and preprint trends, pairwise correlations were computed using three methods: Pearson’s $r$, Spearman’s rank correlation, and Kendall’s tau. These were applied to both the raw annual series and the series with COVID-19-related retractions removed, to distinguish inertial coupling from episodic effects. The resulting coefficients and significance levels are reported in \texttt{correlations\_summary.csv} and are the basis for the interpretation of system-level synchrony discussed in this work.

All scripts, raw data, fitted results, and analysis routines are openly available and fully reproducible.

\subsection*{Limitations}

While this study leverages the largest open-access metadata corpus (CrossRef) and uses multiple regression models, several limitations warrant mention. First, metadata completeness varies across publishers and years, particularly for older retractions and preprints. Second, reliance on title-based filters (e.g., “Retraction Notice”) may omit items with non-standard labeling. Third, the logistic models assume capacity limits without direct evidence of systemic saturation. Finally, while model forecasts diverge beyond 2024, these projections should be interpreted with caution and validated against post-2025 empirical data as it becomes available.

\section*{Code and Data Availability}

All code (including full API queries, analysis scripts, and plotting routines), as well as the underlying time series and regression results, are openly available at \url{https://github.com/khalid-saqr/CrossRef}. The repository includes the files \texttt{analysis\_time\_series.csv} and \texttt{regression\_fits\_detailed.csv} for full reproducibility.

\bibliographystyle{abbrvnat}
\bibliography{references}

\end{document}